\documentclass[aps,preprint,floatfix,altaffilletter]{revtex4}
\usepackage{dcolumn}
\usepackage{color}
\usepackage{amsmath}
\usepackage{amssymb}

\usepackage[dvips]{graphicx}

\newcommand{\be}{\begin{equation}}
\newcommand{\ee}{\end{equation}}
\newcommand{\bea}{\begin{eqnarray}}
\newcommand{\eea}{\end{eqnarray}}

\newcounter{Fig}

\begin{document}

\begin{sloppy}

\title{Nonreciprocal rotating power flow within plasmonic nanostructures}

\author{Arthur R. Davoyan}
\email{davoyan@seas.upenn.edu}

\author{Nader Engheta}
\email{engheta@ee.upenn.edu}

\affiliation{Department of Electrical and Systems Engineering, University of Pennsylvania, Philadelphia,
Pennsylvania 19104, USA }

\date{\today}


\begin{abstract}
We theoretically explore the notion of nonreciprocal near-zone manipulation of electromagnetic fields within subwavelength plasmonic nanostructures embedded in magneto-optical materials.  We derive an analytical model predicting a strong, magneto-optically induced time-reversal symmetry breaking of localized plasmonic resonances in topologically symmetric structures. Our numerical simulations of plasmon excitations reveal a considerable near-zone power flow rotation within such hybrid nanostructures, demonstrating nanoscale nonreciprocity.  This can be considered as another mechanism for tuning plasmonic phenomena at the nanoscale.
\end{abstract}

\pacs{73.20.Mf,78.20.Ls,42.25.-p,78.67.Uh}
\maketitle

Recent advances in plasmonics -- nanoscale electrodynamics of metals -- have provided roadmaps for significant miniaturization of optical devices and components. Consequently, electromagnetic fields can be confined, enhanced, and manipulated in length scales as low as few nanometers around plasmonic structures~\cite{Maier_rev}. Such a dramatic field confinement led to a number of breakthroughs reported lately, including nano-lasers~\cite{Shalaev_laser}, medical treatment~\cite{Halas_cancer}, subwavelength nano-antennas~\cite{Novotny_antenna,Gissen_Yagi}, observation of novel magnetic resonances~\cite{Lukyanchuk_maglight,Nordlander_oligomers} and some other applications~\cite{Brongersma}. However, further scientific development and technological expansion in the fields of plasmonic optics and nanophotonics may in general be accelerated by the ability to tune actively the optical response of plasmonic strutures and their field distributions at the nanoscale. Several techniques developed in recent years such as active all-optical control~\cite{Zayats,active_1,active_2} , nonlinear self-tuning~\cite{Zheludev_nln,Davoyan_taper}, waveform shaping~\cite{Quidant,Minovich,Capasso} and structural modifications~\cite{Engheta_nature,Nordlander_oligomers,Giessen_ruler} are regarded as possible pathways for the nanoscale field manipulation. In particular, in Refs.~\cite{Nordlander_oligomers,Giessen_ruler} a structural design and arrangement of plasmonic oligomers was exploited for precise control over the resonances and corresponding near field distributions.

Mixing magneto-optical (MO) materials with plasmonic structures may provide another mechanism, and an additional degree of freedom, in tailoring the light-matter interaction in the vicinity of plasmonic structures. It is well known that the MO materials may possess {\it nonreciprocal response} as a result of breaking the time-reversal symmetry in optical phenomena~\cite{Soljacic_oneway,Fan_oneway,Ross_isolator}, - a property that may be exploited for signal handling and manipulation, and for photonic circuits~\cite{Ross_isolator,Soljacic_circ,Fan_circ_v1,Fan_circ_v2}. However, the magneto-optical response of naturally occurring materials is usually very weak~\cite{Landau}, and may not be adequate for some potential applications. Recently an incorporation of MO materials with plasmonic structures has gained a significant attention, due to possible enhancement of magneto-optical activity in highly localized fields~\cite{Wang_coreshell,Beloletov_TKMOE}. Enhancement of such {\it macroscopic} (i.e. far field) nonreciprocal effects as Kerr and Faraday rotations~\cite{Wang_coreshell,Jain_coreshell,Tomita_polarKerr,Garcia_nanodisk}, and transverse Kerr effect~\cite{Beloletov_TKMOE,Fedyanin_TKMOE} have been demonstrated. However, to the best of our knowledge, the {\it microscopic} (i.e. near-field) nonreciprocal optical response and its enhancement have not received as much attention.

\begin{figure}
  \begin{center}
      \includegraphics[width=1\columnwidth]{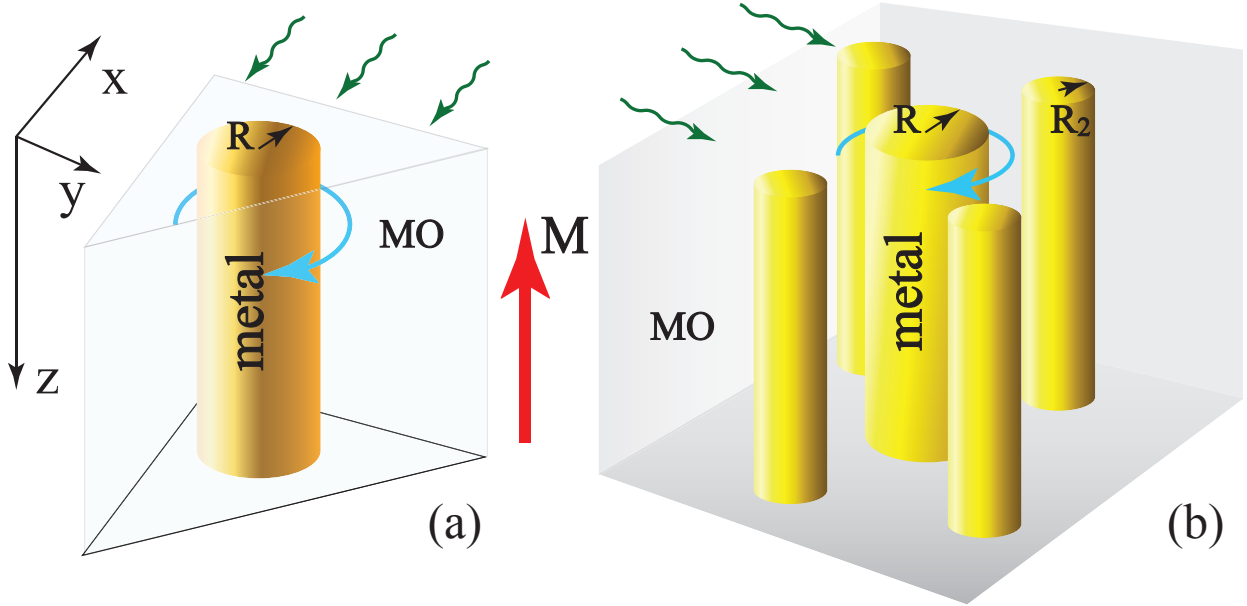}
      \caption{(color online) Schematic of the two structures studied in the Letter: a) A plasmonic nanorod in an equilateral triangular magneto-optical cavity surrounded by a free-space; and b) a collection of nanorods embedded in an unbounded MO medium. Magnetization field shown by a red arrow is directed along the main cylinder axis.} \label{geometry}
  \end{center}
\end{figure}

In this Letter, we theoretically explore and analyze the concept of nonreciprocal manipulation of near-zone optical field and nanoscale power flux by mixing magneto-optical materials with plasmonic nanostructures, and demonstrate, using numerical simulations, a significant enhancement of the nanoscale nonreciprocal response. In particular, we study field distributions at plasmonic resonances in such structures and show that magneto-optical activity may lead to a strong coupling between the degenerate eigenstates corresponding to the same resonant frequency, resulting in the formation of highly localized rotating eigenstates. We analyze the plane wave excitation of corresponding solutions that depend on the strength of magneto-optical activity, and reveal a pronounced power flux circulation. We discuss typical structures where the boosting of near-field power flow circulation is observed, and give a brief insight to the possible experimental realization and potential applications of such observed effect.

We begin our study with the a generalized two-dimensional (2D) eigenfrequency analysis for an arbitrary 2D plasmonic (i.e. metallic) structure embedded in a magneto-optical material. Fig.~\ref{geometry} shows geometries of the problem with the z axis of a Cartesian coordinate system being parallel with the axis of geometries. The magnetization of the MO material may be parallel or anti-parallel with the z axis (i.e. the Voigt configuration). We search for TM electromagnetic field distributions in the $(x-y)$ plane, for which the electric field has $x$ and $y$ components and magnetic field only $z$ component. In this case the MO material response is given by an antisymmetric relative permittivity tensor~\cite{Landau}:

 \begin{equation}
  \bar{\bar{\varepsilon}}= \left( \begin{array}{ccc}
   \varepsilon_{mo} & i\alpha & 0 \\
   -i\alpha & \varepsilon_{mo} & 0 \\
   0 & 0 & \varepsilon_\bot \end{array} \right),
\end{equation}
where $\varepsilon_{mo}$ and $\varepsilon_\bot$ are diagonal components of dielectric permittivity, $\alpha$ is the off-diagonal component of the permittivity tensor responsible for the ``strength'' of magneto-optical activity of the media. Typically $\alpha$ at optical frequencies is very small and usually is of the order of $10^{-2}$ or even smaller. The wave equation for the electric field in the $(x-y)$ plane can be written in the following operator form:

\begin{gather}
  \frac{1}{\varepsilon(x,y)}\left(\begin{array}{cc}
    -\frac{\partial^2}{\partial y^2} & \frac{\partial^2}{\partial x\partial y}\\
    \frac{\partial^2}{\partial x\partial y} & -\frac{\partial^2}{\partial x^2}
  \end{array}\right)
  \left(\begin{array}{c}
   E_x \\ E_y
  \end{array}\right)
  =
  \frac{\omega^2}{c^2}
  \left(\begin{array}{c}
   E_x \\ E_y
  \end{array}\right)+ \nonumber \\
  +i \frac{\alpha}{\varepsilon(x,y)}
  \left(\begin{array}{cc}
   0 & \zeta(x,y)\\
    -\zeta(x,y) & 0
  \end{array}\right)
  \left(\begin{array}{c}
   E_x \\ E_y
  \end{array}\right) \label{wave},
\end{gather}
where $E_x$ and $E_y$ are the electric field components, $\omega$ is the radian frequency, $c$ is the free-space speed of light, $\varepsilon(x,y) = \varepsilon_{mo}$ in MO material and $\varepsilon(x,y) = \varepsilon_m$ in metal, and $\zeta(x,y) = 1$ in MO media and $\zeta(x,y) = 0$ outside of it. Obviously, last term in Eq.(\ref{wave}) associated with the MO activity, leads to a nonreciprocal coupling between the $E_x$ and $E_y$ electric field components. When $\alpha=0$ the Eq.(\ref{wave}) reduces to an ordinary wave equation. Solution of an eigenfrequency problem in a nonmagnetized case yields a set of resonant frequencies $\omega_n^0$ with corresponding set of eigenstates $(E_{x,(n,m)}^0,E_{y,(n,m)}^0)^T$ where the parameters $n$ and $m$ are the modal parameters related to the radial and azimuthal variations of the mode, respectively. In general, the mode degeneracy may be present, implying that $d$ $(m=1..d)$ eigenmodes may possess the same given resonant frequency $\omega_n^0$. Note that, for a nondegenerate system the magneto-optical activity causes the interaction between eigenmodes corresponding to different resonant frequencies $\omega_n^0$, however, when these resonances are well pronounced and well separated from each other such intermode interaction can be considered negligible. At the same time the maximum magneto-optical interaction is expected between the degenerate states corresponding to the same resonance, and in further analysis we focus on this case only. Considering that $\alpha<<\varepsilon_{mo}$, we apply the perturbation method for general case of $d$ times mode degeneracy at a a given resonance $\omega_n^0$ and study the interaction between these degenrate modes. We consider the series expansion of the eigenfrequency and corresponding eigenmodes in terms of powers of $\alpha$ as: $(\omega_n)^2=(\omega_n^0)^2+\alpha(\omega_n^1)^2+{\it O}^2$ and $E_{\nu,n}=\sum c_m E_{\nu,(n,m)}^0+\alpha E_{\nu,n}^1+{\it O}^2$, where $\nu$ stands for the coordinate $x$ or $y$, and $c_m$'s are unknown complex coefficients. Substituting these relations in Eq.(\ref{wave}) and applying standard methods of perturbation theory, we obtain:

\begin{gather}
   \sum \int dS \left[ \left(\frac{\omega_n^1}{c}\right)^2 (\psi_{n,k}^0)^*(\psi_{n,m}^0)+\right. \nonumber \\
   \left.+ i \left(\frac{\omega_n^0}{c}\right)^2 (\psi_{n,k}^0)^* \frac{1}{\varepsilon} \left( \begin{array}{cc} 0&\zeta\\-\zeta&0 \end{array}\right) \psi_{n,m}^0 \right] c_m =0 \label{matrix}
\end{gather}
where the summation is taken over the index $m$ only ($m=1..d$), $\psi_{n,m}^0=(E_{x,(n,m)}^0,E_{y,(n,m)}^0)^T$, and $\omega_n^1$ is the first order correction to the resonant frequency $\omega_n^0$ due to the presence of MO activity, i.e. when $\alpha$ is non-zero.

\begin{figure}[t]
  \begin{center}
      \includegraphics[width=1\columnwidth]{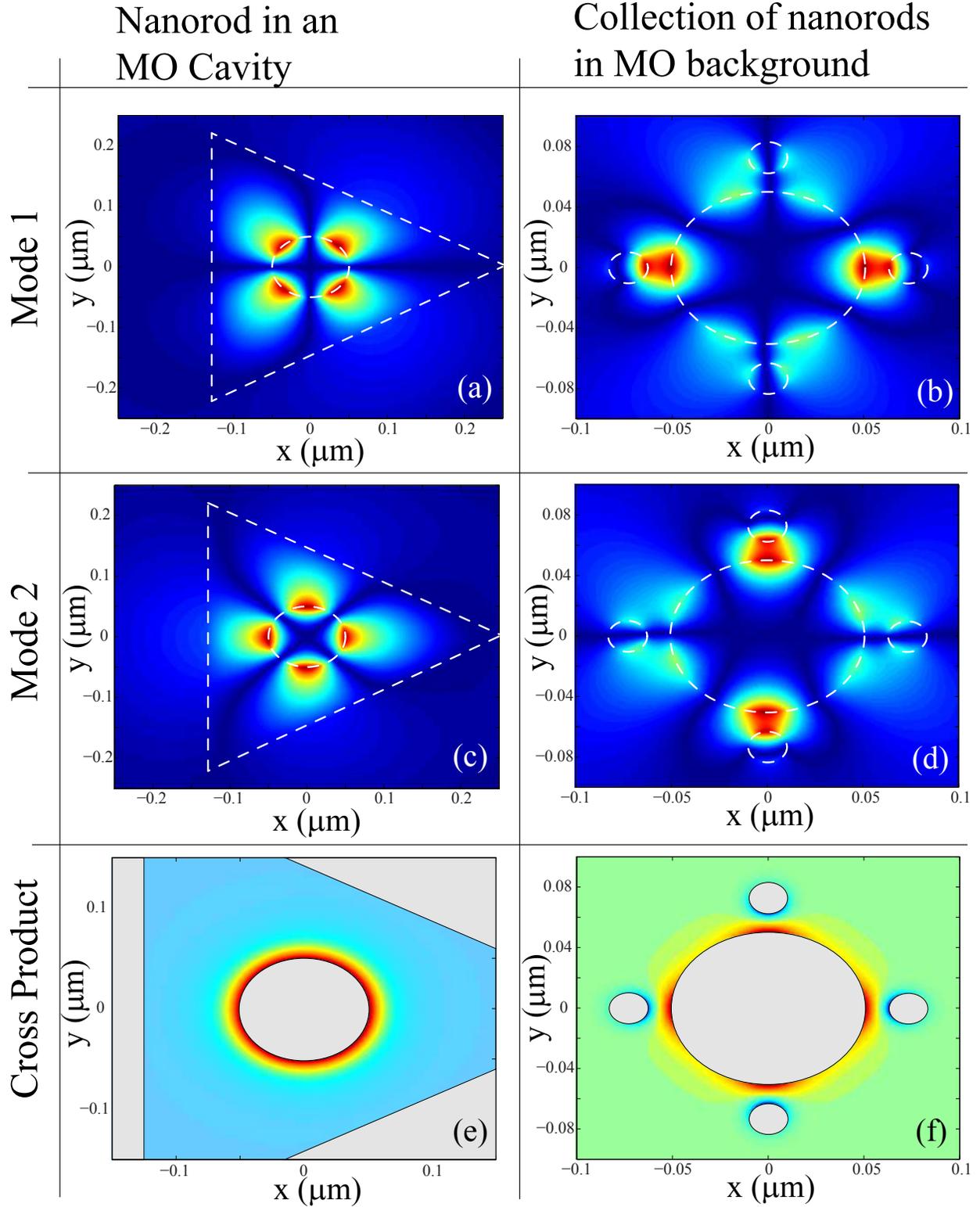}
      \caption{(color online) Intensity profiles of magnetic field $|H|^2$ for the degenerate states in the two structures at their second order resonance ($n=2$) and cross-product between these states for a nanorod in the MO cavity (a,c,e) at $\omega_0=404$THz, and the collection of nanorods embedded in the MO medium at $\omega_0=336$THz (b,d,f), respectively.}\label{fields}
  \end{center}
\end{figure}

Assuming an idealized lossless system, we obtain real valued solutions $\psi_{n,k}^0$ in the zero-th order approximation (i.e $\alpha=0$). In this case it is possible to show that the last term under the integral in Eq.(\ref{matrix}) vanishes for $m = k$, whereas the first term being the scalar product of the eigen-modes is zero for $m\neq k$ due to the orthogonality. Taking these into account one can show that Eq.(\ref{matrix}) reduces to a eignevalue problem for a $d\times d$ matrix with $c_m$ being the eignevectors of this matrix and $\omega_n^1$ unknown eigenvalues. The solutions of such matrix are defined by its zero determinant. 
Eq.(\ref{matrix}) shows that MO activity leads to the coupling between the degenerate eigenstates, which depends on the cross-product between the $k$ and $m$ states, i.e. the last term in Eq.(\ref{matrix}), and results in the formation of a novel set of complex eigenstates $\sum c_m^l \psi_{n,m}^0$ corresponding to the $l$-th eigenvalue of Eq.(\ref{matrix}). The physical meaning of such coupling is that the back-and-forth energy beating between the $k$-th and $m$-th states is established.

For a doubly degenerate case the determinant of Eq.(\ref{matrix}) is written in a simple way:
\begin{gather}
  \left(\omega_n^1\right)^4 < (\psi_{n,1}^0)^2> < (\psi_{n,2}^0)^2> - \nonumber \\
  -\left(\omega_n^0\right)^4 \left[< \frac{1}{\varepsilon}\psi_{n,1}^0 \left( \begin{array}{cc} 0&\zeta\\-\zeta&0 \end{array}\right)\psi_{n,2}^0>\right]^2=0,
\end{gather}
where $<\bullet>=\int \bullet dS$. Considering that $< (\psi_{n,1}^0)^2> \simeq < (\psi_{n,2}^0)^2>$ we derive the expression for the frequency splitting:

\begin{gather}
  \omega_{\pm} \simeq \omega^0 \left(1\pm\frac{\alpha}{\varepsilon_{mo}}\frac{<(E_{x1}E_{y2}-E_{y1}E_{x2})>_{|_{mo}}}{2<(E_{x1})^2+(E_{y1})^2>}\right) \label{split}
\end{gather}
here we have dropped indexes $n$ and $0$ for simplicity. In this case the eigenvectors of Eq.(\ref{matrix}) for $d=2$ are complex conjugates, which means that the solutions of Eq.(\ref{matrix}) to the fist order approximation are ${\bf E}_{\pm}=(E_{x1}\pm i E_{x2},E_{y1}\pm i E_{y2})$ with corresponding eigenfrequencies $\omega_{\pm}$. In analogy with clockwise and counterclockwise plasmonic states of a single nanorod~\cite{Luk_scat} these new solutions $E_\pm$ can be classified as counter-rotating eigenstates. Note that similar solutions were obtained previously for a photonic crystal circulator~\cite{Fan_circ_v1,Fan_circ_v2}. When the frequency split $|\omega_+-\omega_-|$ is relatively large, these states can not be excited with the same strength by an incident electromagnetic field, and therefore the superpositions of these states gives rise to a near-zone field with a rotating near-field pattern. It is important to note here that for doubly degenerate structure possesing a 90 degree rotational symmetry between the eigenmodes (simplest example of such a structure would be a single plasmonic nanorod surrounded by MO) the cross product $<(E_{x1}E_{y2}-E_{y1}E_{x2})>_{|_{mo}}$ vanishes, implying that the reported mechanism of mode coupling and corresponding symmetry breaking do not take place.

The specific type of modal degeneracy depends on the structure design. Generally the degeneracy is observed for structures with some rotational symmetry. As two examples, we consider here a 2D plasmonic nanorod in a MO 2D cavity surrounded by air, and a collection of parallel nanorods immersed in a MO background (see Fig.~\ref{geometry}). In particular, we study a plasmonic nanorod with radius $R=50$nm centered in a MO equilateral triangular cavity with side $450$nm, surrounded by air, Fig.~\ref{geometry}(a), and a plasmonic nanorod with radius $R=50$nm surrounded by four other nanorods with smaller radii $R_2 =10$nm symmetrically centered around the first nanorod, $80$nm away from its axis, Fig.~\ref{geometry}(b). Without loss of generality, here we assume that the lossless metal dielectric constant is fixed at $\varepsilon_m=-10$, and we consider our MO material to be a Bismuth Iron Garnet (BIG) with  $\varepsilon_{mo}=6.25$ and $\alpha$ as a free parameter for our parametric study.

\begin{figure}[h]
  \begin{center}
      \includegraphics[width=1\columnwidth]{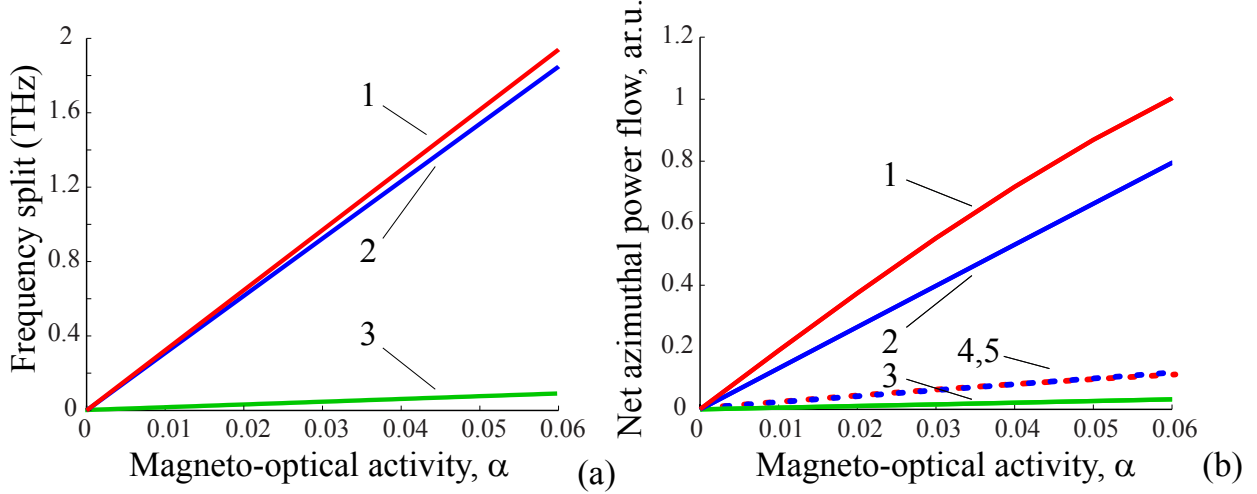}
      \caption{(color online) (a) Frequency splitting $|\omega_+-\omega_-|$ vs magneto-optical activity parameter $\alpha$ and (b) net azimuthal power flow around the main cylinder, $<P_{\phi}>$. Curves 1-3 correspond to the nanorod in the MO cavity, the collection of nanorods embedded in MO medium, and the single nanorod surrounded by uniform MO, respectively. Curves 4,5 the same as 1,2 but for a lossy nanorods.}\label{charact}
  \end{center}
\end{figure}

First, we consider no MO activity (i.e. $\alpha=0$, for zero magnetization) and search for the eigenfrequencies and corresponding modal distributions using numerical simulation with COMSOL Multiphysics, a commercially available finite-element-based simulation software. The eigenfrequency analysis shows that both structures possess two doubly degenerate modes at the second order plasmonic resonance ($n=2$). The corresponding distribution of the magnetic field intensities for both degenerate states are shown in Figs.~\ref{fields}(a,c,b,d). One can see that the fields are highly localized in an area less than $100\times100$nm$^2$. The cross-products of these degenerate modes are shown in Figs.~\ref{fields}(e,f). Clearly, the cross-product is azimuthally  symmetric and monotonically decaying along the radial direction, implying that overall integral $<(E_{x1}E_{y2}-E_{y1}E_{x2})>_{|_{mo}}$ would be nonzero and therefore a strong magneto-optical coupling will exist between the corresponding modes, if the MO activity is introduced. When the structure is magnetized, i.e $\alpha\neq0$, the two counter-rotating states with $\omega_{\pm}$ eigenfrequencies are formed, as was mentioned above. In Fig.~\ref{charact}(a) we plot the values of the frequency split $|\omega_+-\omega_-|$ as a function of MO parameter $\alpha$, using Eq.(\ref{split}). We note that even for small values of $\alpha$, the frequency split is significant and is up-to $10$ times higher than that reported in Ref.~\cite{Fan_circ_v1}. Noticeably the coupling strength for both structures studied in this Letter is almost analogous, which can be attributed to the similarities in the topology of near-field distributions, see Fig.~\ref{fields}. It is important to compare the reported mechanism of symmetry breaking with that of single plasmonic nanorod ($R=50nm$) surrounded by magneto-optical media. Although the nanorod has two degenerate states at each plasmonic resonance ($n=1,2,...$) the cross-product between them is zero, implying that the reported mechanism of symmetry breaking is not applicable for such nanorod so that the corresponding frequency splitting is two orders of magnitude smaller, Fig.3(a) (note that the MO activity still leads to the symmetry breaking between these degenerate modes associated with an interface effect~\cite{Camley}).

Next, we study the excitation of the predicted rotating states. In each of the geometries studied here, we consider a plane wave incident on the structure, as schematically shown in Fig.~\ref{geometry}. We numerically simulate this problem using COMSOL Multiphysics software. We trace the near-field distributions at these observed resonances and analyze their variation as a function of the MO parameter $\alpha$. In Fig.~\ref{circulation} we present distributions of magnetic field and the Poynting vector for a nanorod in a triangular MO cavity at its second order plasmonic resonance ($n=2$). We observe a strong distortion and one way-rotation of the near-field pattern and its power flux, when MO activity is present. The Poynting vector, describing local power flow, experiences a dramatic change in its behavior as one increases $\alpha$. In particular, for an $\alpha=0$ case we observe a mirror-symmetric power flow around the nanorod, Fig.~\ref{circulation}(a), i.e. there is no preferred direction of rotation for this power flow. However, when the MO parameter $\alpha$ is non zero, we note a strong power flow circulation around the nanorod as shown in Fig.~\ref{circulation}(b). Similar behavior is found for the structure shown in Fig.~\ref{geometry}(b), where again the magneto-optical activity leads to the breaking of mirror symmetry of power flow around the structure (not shown here). We note that the analysis of field distributions at the first order plasmonic resonances ($n=1$) does not reveal any pronounced energy circulation, although both geometries are doubly degenerate at this resonance as well. The latter is due to the zero cross-product between the modes.

In order to quantitatively determine the strength of such Poynting vector circulation we evaluate the net azimuthal power flow by inscribing the structure into a mathematical circular region and then we calculate $<P_{\phi}>=|\int\limits_0^L\int\limits_0^{2\pi}\zeta\overline{P}d\overline{\phi}dr|$, where $\overline{\phi}$ is the azimuthal unit vector, $\overline{P}$ is the Poynting vector, and $L=150$nm is the radius of our integration domain arbitrarily selected. In Fig.~\ref{charact}(b), we plot this value for both of our structures. We observe almost linear increase in the net power flow circulation with the increase of MO activity. Clearly, the rotational power flow as a function of $\alpha$ correlates well with the frequency split vs $\alpha$ shown in Fig.~\ref{charact}(a). Owing to the nonreciprocal nature of the problem, the rotational direction and net power flow of circulation are dependent on  the direction of the magnetization. We also study the plane wave excitation of a single nanorod surrounded by a magneto-optical medium, and analyze the corresponding net azimuthal power flow circulation, see Fig.~\ref{charact}(b). As can be seen, the overall circulation in this case is almost two orders of magnitude smaller, ensuring that the near-field rotation is negligibly weak, which agrees well with frequency splitting predictions shown in Fig.~\ref{charact}(a).

\begin{figure}[t]
  \begin{center}
      \includegraphics[width=1\columnwidth]{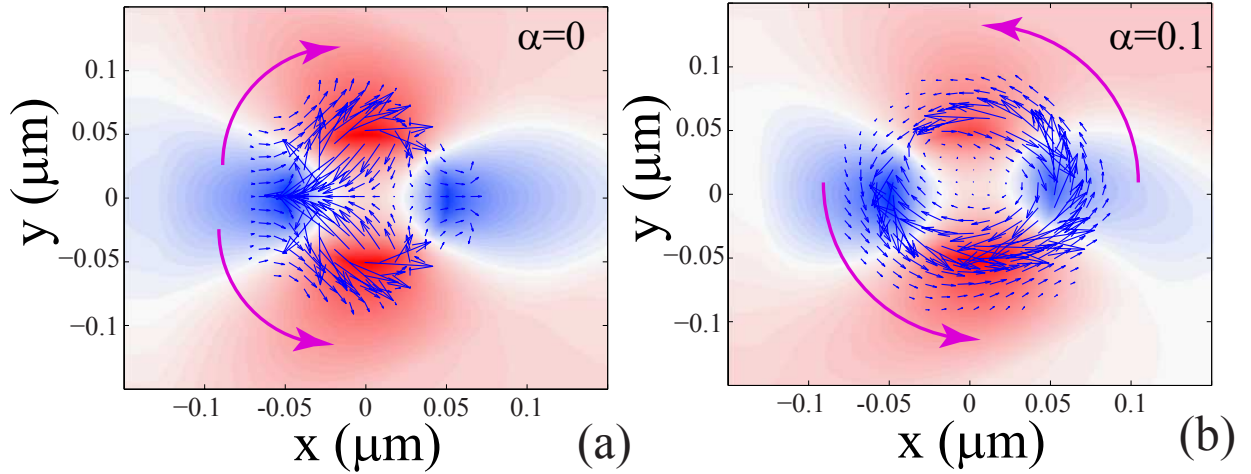}
      \caption{(color online) Magnetic field (colormap) and Poynting vector (arrows) distributions for a plasmonic nanorod in a triangular MO cavity at the second order plasmonic resonance ($\omega=404$THz) when $\alpha=0$ (a) and $\alpha=0.1$ (b).} \label{circulation}
  \end{center}
\end{figure}

Finally, we study the influence of losses on the system dynamics. In particular, we trace the power flux circulation in a lossy system taking into account the metal loss, i.e. considering that $\textrm{Im}(\varepsilon_m) = 0.1$. Corresponding net azimuthal power flow is shown as dashed curves in Fig.~\ref{charact}(b). Evidently, the nanocirculation of the power flux around the plasmonic nanostructures is well preserved in these lossy systems, and is one order of magnitude higher than that for a single lossless cylinder in unbounded MO medium. The predicted phenomenon of nanoscale circulation of power flux due to combination of the plasmonics and MO activity may have interesting impacts on various scenarios ranging from optical tweezing and nanoscale paticle manipulation and rotation, to nanoantennas, and nanocircuits.

In conclusion, using analytical and numerical study we have shown that incorporating magneto-optical materials with plasmonic structures leads to a significant enhancement of near-field nonreciprocal response. We have demonstrated that in the presence of MO activity the modes excited at plasmonic resonances in certain types of geometries couple strongly with each other resulting in a formation of nonreciprocal rotating states. The plane wave excitation of two plasmonic geometries studied here has revealed a strong nearfield power flow circulation. This near-zone subwavelength power flux circulation and rotation may be exploited for design of next generation tunable nanoscale plasmonic devices. The principles of nanoscale optical field manipulation might pave the way for a wide variety of potential applications in tunable sensing, active plasmonic elements, particle manipulation, and others. This work is supported in part by the US Air Force Office of Scientific Research (AFOSR) grant number  FA9550-10-1-0408.

\end{sloppy}

\begin{thebibliography}{99}

\bibitem{Maier_rev} S.A. Maier, {\em Plasmonics: Fundamentals and Applications} (Springer-Verlag, Berlin 2007).
\bibitem{Shalaev_laser} M.A. Noginov, G. Zhu, A.M. Belgrave, R. Bakker, V.M. Shalaev, E.E. Narimanov, S. Stout, E. Herz, T. Suteewong, and U. Wiesner, Nature Photon. {\bf 460}, 1110-1112 (2009).
\bibitem{Halas_cancer} S. Lal, S.E. Clare, and N.J. Halas, Acc. Chem. Res. {\bf 41}, 1842-1851 (2008).
\bibitem{Novotny_antenna} L. Novotny and N.F. van Hulst, Nature Photon. {\bf 5}, 83-90 (2011).
\bibitem{Gissen_Yagi} D. Dregely, R. Taubert, J. Dorfmuller, R. Vogelgesang, K. Kern, and H. Giessen, Nature Commun. {\bf 2}, 267 (2011).
\bibitem{Lukyanchuk_maglight} A.I. Kuznetsov, A.E. Miroshnichenko, Y.H. Fu, J.B. Zhang, and B. Luk'yanchuk, Scientific Rep. {\bf 2}, 492 (2012).
\bibitem{Nordlander_oligomers} J. Ye, F. Wen, H. Sobhani, J.B. Lassiter, P. van Dorpe, P. Nordlander, N.J. Halas, Nano Lett. {\bf 12}, 1660-1667 (2012).
\bibitem{Brongersma} { Surface Plasmon Nanophotonics: Springer series in optical sciences} M.L. Brongersma and P.G. Kik, Eds. (Springer-Verlag New York, LLC
, 2007).

\bibitem{Zayats} M. Kauranen, and A.V. Zayats, Nat. Photon. {\bf 6}, 737, (2012).
\bibitem{active_1} K.F. MacDonald, Z.L. Samson, M.I. Stockman, and N. I. Zheludev, Nature Photon. {\bf 3}, 55 (2008).
\bibitem{active_2} S. Papaioannou,	 D. Kalavrouziotis,	 K. Vyrsokinos,	 J.-C. Weeber,	 K. Hassan,	 L. Markey,	 A. Dereux,	 A. Kumar,	 S.I. Bozhevolnyi,	 M. Baus,	 T. Tekin,	 Di. Apostolopoulos,	 H. Avramopoulos, and N. Pleros, Scientific Rep. {\bf 2}, 652 (2012).

\bibitem{Zheludev_nln} M. Ren, E. Plum, J. Xu, and N.I. Zheludev, Nature Commun. {\bf 3}, 833 (2012).
\bibitem{Davoyan_taper} A.R. Davoyan, I.V. Shadrivov, A.A. Zharov, D.K. Gramotnev, and Y.S. Kivshar, Phys. Rev. Lett. {\bf 105}, 116804 (2010).

\bibitem{Quidant}  G. Volpe, S. Cherukulappurath, R.J. Parramon, G. Molina-Terriza, and R. Quidant, Nanolett. {\bf 9}, 3608-3611 (2009).
\bibitem{Minovich}A. Minovich, A.E. Klein, N. Janunts, T. Pertsch, D.N. Neshev, and Y.S. Kivshar, Phys. Rev. Lett. {\bf 107}, 116802 (2011).
\bibitem{Capasso} J. Lin, J. Dellinger, P. Genevet, B. Cluzel, F. de Fornel, and F. Capasso, Phys. Rev. Lett. {\bf 109}, 093904 (2012).

\bibitem{Engheta_nature} Y. Sun, B. Edwards, A. Alu, and N. Engheta, Nature. Mater. {\bf 11}, 208-212 (2012).
\bibitem{Giessen_ruler} N. Liu, M. Hentschel, T. Weiss, A.P. Alivisatos, and H. Giessen, Science {\bf 332},  1407-1410 (2011).


\bibitem{Soljacic_oneway} Z. Wang, Y. Chong, J.D. Joannopoulos, and Marin Soljacic, Nature {\bf 461}, 772-775 (2009).
\bibitem{Fan_oneway} Z. Yu, G. Veronis, Z. Wang, and S. Fan, Phys. Rev. Lett. {\bf 100}, 023902 (2008).
\bibitem{Ross_isolator} L. Bi, J. Hu, P. Jiang, D.H. Kim, G.F. Dionne, L.C. Kimerling, and C.A. Ross, Nature Photon. {\bf 5}, 758-762 (2011).

\bibitem{Soljacic_circ}  W. Qiu, Z. Wang, and M. Soljacic, Opt. Express {\bf 19}, 22248-22257 (2011).
\bibitem{Fan_circ_v1} Z. Wang and S. Fan, Opt. Lett. {\bf 30}, 1989-1991 (2005).
\bibitem{Fan_circ_v2} S. Fan and Z. Wang, J. Magn. Soc. Jpn. {\bf 30}, 641-645 (2006).

\bibitem{Landau}L D Landau, L. P. Pitaevskii, E.M. Lifshitz, {\em Electrodynamics of Continuous Media, Second Edition: Volume 8 (Course of Theoretical Physics)} (Butterworth-Heinemann, 1984).

\bibitem{Wang_coreshell} L. Wang, C. Clavero, Z. Huba, K.J. Carroll, E.E. Carpenter, D. Gu, and R.A. Lukaszew, Nano Lett. {\bf 11}, 1237-1240 (2011).
\bibitem{Beloletov_TKMOE} V.I. Belotelov, I.A. Akimov, M. Pohl, V.A. Kotov, S. Kasture, A.S. Vengurlekar, A.V. Gopal, D.R. Yakovlev, A.K. Zvezdin, and M. Bayer, Nature Nanotech. {\bf 6}, 370-376 (2011).
\bibitem{Jain_coreshell} P.K. Jain, Y. Xiao, R. Walsworth, and A.E. Cohen, Nano Lett. {\bf 9}, 1644-1650 (2009).
\bibitem{Tomita_polarKerr} S. Tomita, T. Kato, S. Tsunashima, S. Iwata, M. Fujii, and S. Hayashi, Phys. Rev. Lett. {\bf 96}, 167402 (2006).
\bibitem{Garcia_nanodisk} B. Sepulveda, J.B. Gonzalez-Diaz,  A. Garcia-Martin, L.M. Lechuga, and G. Armelles, Phys. Rev. Lett. {\bf 104}, 147401 (2010).
\bibitem{Fedyanin_TKMOE} A. A. Grunin, A. G. Zhdanov, A. A. Ezhov, E. A. Ganshina, and A. A. Fedyanin, Appl. Phys. Lett. {\bf 97}, 261908 (2010).

\bibitem{Luk_scat} B.S. Lukyanchuk and V. Ternovsky, Phys. Rev. B {\bf 73}, 235432 (2006).
\bibitem{Camley} R.E. Camley, Surf. Science Rep. {\bf 7}, 103-187 (1987).
\end{thebibliography}
\end{document}